\input ./style/arxiv-general.cfg
\documentclass[aoas,MSNbibl,nameyear,seceqn,dvips]{arximspdf}
\makeatletter
   \@ifpackageloaded{graphicx}{}{\usepackage{graphicx}}
\makeatother
\usepackage{dcolumn}

\doi{10.1214/15-AOAS810}% Updated by VTEXPTS2LaTeX.exe, 30.03.2015 09:53
\volume{9}
\issue{2}
\pubyear{2015}
\firstpage{714}
\lastpage{730}
\docsubty{FLA}

\makeatletter
\newcolumntype{d}[1]{D{.}{.}{#1}}
\renewcommand{\mid}{|}
\newcommand{\overset}{\stackrel}
\makeatother

\begin{document}
\begin{frontmatter}

%\dochead{}
\title{Semiparametric time to event models in the presence of error-prone, self-reported outcomes---With application to the~women's~health~initiative\thanksref{T11}}
\runtitle{Time to event models for self-reported outcomes}

\begin{aug}
% Corresponding author: Xiangdong Gu - ustcgxd@gmail.com% Updated by
%VTEXPTS2LaTeX.exe, 30.03.2015 09:53
\author[A]{\fnms{Xiangdong}~\snm{Gu}\corref{}\thanksref{M1}\ead[label=e1]{ustcgxd@gmail.com}},
\author[B]{\fnms{Yunsheng}~\snm{Ma}\thanksref{M2}\ead[label=e2]{Yunsheng.Ma@umassmed.edu}}
\and
\author[A]{\fnms{Raji}~\snm{Balasubramanian}\thanksref{M1,T22}\ead[label=e3]{rbalasub@schoolph.umass.edu}}
\runauthor{X. Gu, Y. Ma and R. Balasubramanian}
\affiliation{University of Massachusetts Amherst\thanksmark{M1}
and\break  University of Massachusetts Medical School\thanksmark{M2}}
%\dedicated{}
\address[A]{X. Gu\\
R. Balasubramanian\\
Department of Biostatistics\\
\quad and Epidemiology\\
University of Massachusetts\\
Amherst, Massachusetts 01003\\
USA\\
\printead{e1}\\
\phantom{E-mail: }\printead*{e3}}
\address[B]{Y. Ma\\
Department of Medicine\\
Division of Preventive\\
\quad and Behavioral Medicine\\
University of Massachusetts Medical School\\
Worcester, Massachusetts 01655\\
USA\\
\printead{e2}}
\end{aug}
\thankstext{T11}{The WHI program is funded by the National Heart, Lung, and Blood
Institute, National Institutes of Health, U.S. Department of Health and
Human Services through contracts HHSN268201100046C, HHSN268201100001C,
HHSN268201100002C, HHSN268201100003C, HHSN268201100004C and HHSN271201100004C.}
\thankstext{T22}{Supported by the National Institutes of Health 1R01HL122241-01A1.}

% HISTORY:
%
\received{\smonth{9} \syear{2014}}% Updated by VTEXPTS2LaTeX.exe,
%30.03.2015 09:53
%
\revised{\smonth{1} \syear{2015}}% Updated by VTEXPTS2LaTeX.exe,
%30.03.2015 09:53

% ABSTRACT
%
\begin{abstract}
The onset of several silent, chronic diseases such as diabetes can be
detected only through
diagnostic tests. Due to cost considerations, self-reported outcomes
are routinely collected
in lieu of expensive diagnostic tests in large-scale prospective
investigations such as the
Women's Health Initiative. However, self-reported outcomes are
subject to imperfect
sensitivity and specificity. Using a semiparametric likelihood-based approach,
we present time to event models to estimate the association of one or
more covariates with
a error-prone, self-reported outcome. We present simulation studies to
assess the effect of error in self-reported outcomes with regard to
bias in the estimation of the regression parameter of interest. We
apply the proposed methods to prospective data from 152,830 women
enrolled in the Women's Health Initiative to evaluate the effect of
statin use with the risk of incident diabetes mellitus among
postmenopausal women. The current analysis is based on follow-up
through 2010, with a median duration of follow-up of 12.1 years. The
methods proposed in this paper are readily implemented using our freely
available R software package \textit{icensmis}, which is available at the
Comprehensive R Archive Network (CRAN) website.
\end{abstract}

% KEYWORDS
% Pirmas kwd is didziosios raides
%
\begin{keyword}
\kwd{Measurement error}
\kwd{panel data}
\kwd{interval censoring}
\kwd{time to event outcomes}
\end{keyword}
\end{frontmatter}

%s1 #&#
\section{Introduction}
The onset of several chronic diseases such as diabetes are asymptomatic
and can be detected only through diagnostic tests. For example,
diabetes can be detected by measuring levels of fasting blood glucose
or glycosylated hemoglobin levels (HbA1c). However, the costs of such
gold standard diagnostic tests can be prohibitive in large-scale
epidemiological studies such as the Women's Health Initiative (WHI)
that enroll and follow over a hundred thousand subjects. Disease
prevalence and incidence in large observational cohorts are often
ascertained through error-prone, self-reported questionnaires. In this
paper, we propose a semiparametric model to assess the association of
specific covariates of interest with a silent time to event outcome
that is assessed through periodic, error-prone self-reports.

Using data from postmenopausal women enrolled in the WHI, the
motivating application in this paper is the evaluation of the
hypothesis that the use of cholesterol lowering medications (statins)
can result in an increased risk of diabetes. The WHI recruited women
($N={}$161,808) aged 50--79 at 40 clinical centers across the U.S. from
1993--1998 with ongoing follow-up [\citet{whidesign}]. Prevalent
and incident diabetes during the course of follow-up was ascertained by
self-report obtained at each annual visit. In a recent paper,
\citet{ma2012} presented an analysis of the effects of statin use
on the risk of incident diabetes in the WHI using Cox proportional
hazards models. The analyses were conducted based on the assumption
that self-reported outcomes of prevalent and incident diabetes are
error-free. The validity of self-reports of incident and prevalent
diabetes have been evaluated using data from a substudy nested within
the WHI---when compared to fasting glucose levels (treated as the gold
standard), diabetes self-reports had a positive predictive value of
74\% and negative predictive value of 97\% [\citet
{b46,margolis2}]. Other studies such as the Nurses' Health Study,
Physicians' Health Study and the Finnish Public Sector Study also
commonly use self-reported outcomes [\citet{b32,b34,finnish}].

When a perfect diagnostic test is given sequentially at different
points in time to the same individual, the time until the event of
interest can be determined to lie in the interval between the last
negative test and the first positive test---that is, the time until the
event is interval censored. In this context, methods for estimating the
survival distribution and assessing the effect of covariates have been
developed [\citet{b68,b20}]. However, when error-prone diagnostic
procedures such as self-reports are used, standard methods for interval
censored outcomes are rendered invalid. Previous work in this area
includes methods for error-prone outcomes with application to data
collected from laboratory-based diagnostic tests in studies in HIV, HPV
and STD [\citeauthor{b5} (\citeyear{b5,b6}), \citet{b48,b49}]. \citet{b6} developed a
formal likelihood framework to estimate the distribution of the time to
mother to child transmission of HIV. The proposed methods were applied
to data from imperfect DNA PCR diagnostic tests to detect the presence
of HIV in infants who were born to HIV-positive pregnant women.
\citet{b49} extended the discrete proportional hazard model to
incorporate outcomes and covariates. In related work, several papers
proposed generalized Cox models in settings involving time to event
outcomes with incomplete event adjudication [\citet{b60,b14,b15}].
Other related work includes that proposed by \citet{b48} in the
context of HPV studies, where the authors accommodate misclassification
by incorporating ideas of binary generalized linear models with
outcomes subject to misclassification [\citet{b51}]. The problem
of error-prone time to event outcomes can also be handled through the
Hidden Markov Model (HMM) framework. Previous applications of HMM-based
methods include the areas of breast cancer [\citet{b11}], HIV
[\citet{b56,b27}], lung transplantation [\citet{b35}] and
cervical smear tests [\citet{b38}]. \citet{b36} present a
general framework for staged Markov models to handle misclassification
due to error-prone screening tests. Other recent methodological
advances within the general area of outcomes measured with error
include the papers by \citet{garcia-zattera} and \citet
{lyles2011}, as well as works on covariate measurement error with
application to the WHI and the Nurses Health Study [\citet
{prentice2012,spiegelman}]. However, none of the previous literature
specifically considers error-prone, self-reported time to event outcomes.

In this paper we present a likelihood-based approach to incorporate
time-varying covariate effects specific to the setting in which the
prevalence and incidence (time to event) of a chronic condition such as
diabetes is ascertained through error-prone self-reports. We
incorporate the situation where an unknown proportion of subjects who
have already experienced the event of interest at baseline are
mistakenly included into the study, due to the use of error-prone
self-reports at study entry. We also provide a freely available R
software package and illustrate its use [\citet{icensmis}]. In
Section~\ref{s:method} we present notation, form of the likelihood
function, address issues related to estimation and extensions to
incorporate misclassification of subjects at study entry. In
Section~\ref{s:simulation} we perform simulation studies to evaluate
the effects of various degrees of error in self-reports. We investigate
the effects of erroneous inclusion of subjects who have already
experienced the event of interest due to imperfect negative predictive
values associated with self-reports. In Section~\ref{s:application} we
evaluate the association between statin use with the risk of incident
diabetes in a subset of 152,830 women enrolled in the WHI. Last, in
Section~\ref{s:discussion} we discuss the findings of this study and
highlight future directions.

%s2 #&#
\section{Methods}
\label{s:method}
In this section we present notation, form of the likelihood and
extensions to incorporate the possibility of misclassification at study entry.

%s2.1 #&#
\subsection{Notation, likelihood, estimation}
\label{s:likelihood}
Let $X$ refer to the random variable denoting the unobserved time to
event for an individual, with associated survival, density and hazard
functions denoted by $S(x), f(x)$ and $\lambda(x)$, for $x \ge0$,
respectively. The time origin is set to 0, corresponding to the
baseline visit at which all subjects enrolled in the study are
event-free. In other words, $\operatorname{Pr}(X > 0) = 1$. Without loss of
generality, we set $X=\infty$ when the event of interest does not
occur. Let $N$ denote the number of subjects and $n_i$ denote the number
of visits for the $i$th subject. At each visit, we assume that each
subject would self-report their disease status. For example, in the
WHI, information on incident diabetes was collected at periodically
scheduled visits using self-reported questionnaires. For the $i$th
subject, we let $\mathbf{R}_i$ and $\mathbf{t}_i$ denote the $1 \times
n_i$ vectors of self-reported, binary outcomes and corresponding visit
times, respectively. In particular, $R_{ij}$ is equal to 1 if the $j$th
self-report for the $i$th subject is positive (indicating occurrence of
the event of interest such as diabetes) and 0 otherwise. We assume that
self-reports are collected at prescheduled visits up to the time of the
first positive self-report, thus, the vectors of test results ($\mathbf{R}_i$), visit times ($\mathbf{t}_i$) and the number of self-reports
collected per subject ($n_i$) are random. Let $\tau_1, \ldots, \tau_J$
denote the distinct, ordered visit times in the data set among $N$
subjects, where $0=\tau_0<\tau_1<\cdots<\tau_J<\tau_{J+1}=\infty$,
thus, the time axis can be divided into $J+1$ disjoint intervals, $[0,
\tau_1),[\tau_1,\tau_2),\ldots,[\tau_{J},\infty)$.

The joint probability of the observed data for the $i$th subject can be
expressed as
\begin{eqnarray*}
g(\mathbf{R}_i, \mathbf{t}_i, n_i) &=&
\sum_{j = 1}^{J+ 1}{\operatorname{Pr}(\tau_{j-1}
< X_i \le\tau_{j}) \operatorname{Pr}(\mathbf{R}_i,
\mathbf{t}_i, n_i \mid\tau_{j-1} <
X_i \le\tau_{j})}
\\
&=& \sum_{j = 1}^{J+ 1}{ \theta_j
\operatorname{Pr}(\mathbf{R}_i, \mathbf{t}_i, n_i
\mid\tau_{j-1} < X_i \le\tau_{j})},
\end{eqnarray*}
where $\theta_j = \operatorname{Pr}(\tau_{j-1} < X \le\tau_{j})$,
$\tau_0 = 0$ and $\tau_{J+1} = \infty$.

To simplify the form of the expression above, we make the assumption
that given the true time of event $X_i$, an individual's $n_i$
self-reports are independent. That is,
\begin{eqnarray*}
\operatorname{Pr}(\mathbf{R}_i \mid X_i, \mathbf{t}_i)
&=& \prod_{k = 1}^{n_i}{\operatorname{Pr}(r_{ik}
\mid X_i, t_{ik}) }.
\end{eqnarray*}
This assumption implies that the observed values of other self-reported
outcomes do not provide
additional information about the distribution of a particular
self-reported outcome from that
provided by the actual time of the event.

Based on the derivation in \citet{b6}, it can be shown that the
joint probability of the observed data for the $i$th subject can be
simplified as
\begin{eqnarray}\label{eq2.1}
\nonumber
g(\mathbf{R}_i, \mathbf{t}_i,
n_i) &=& \sum_{j = 1}^{J+ 1}{
\theta_j \Biggl[ \prod_{k=1}^{n_i}{
\operatorname{Pr}(r_{ik} \mid\tau_{j-1} < X_i \le
\tau_{j}, t_k) } \Biggr]}
\nonumber\\[-8pt]\\[-8pt]\nonumber
&=& \sum_{j = 1}^{J+ 1}{
\theta_j C_{ij}},
\end{eqnarray}
where $C_{ij} = [ \prod_{k=1}^{n_i}{\operatorname{Pr}(r_{ik} \mid\tau
_{j-1} < X_i \le\tau_{j}, t_k) } ]$.
We assume that the probability of a positive self-report at the $k$th
visit ($r_{ik}=1$) conditional on the interval containing the true
event time and visit time can be expressed as
\[
\operatorname{Pr}(r_{ik}=1 | \tau_{j-1} < X_i \le
\tau_j, t_k)= \cases{ \varphi_1, &\quad
$t_k \ge\tau_{j}$,
\cr
1-\varphi_0, &\quad
$t_k \le\tau_{j-1}$.}
\]
Here, $\varphi_1$ and $\varphi_0$ denote the sensitivity and
specificity of self-reports, respectively. Thus, the terms $C_{ij}$,
for $j=1, \ldots, J+1$, in equation (\ref{eq2.1}) can be expressed as a product
involving the constants $\varphi_1$ and $\varphi_0$. Thus, in the
absence of covariates, the log likelihood for a random sample of $N$
subjects can be expressed as
%
%e2.1 #&#
\begin{equation}\label{eq2.2}
l(\bolds{\theta})=\log\bigl(L(\bolds{\theta})\bigr)=\sum
_{i=1}^N\log\Biggl(\sum
_{j=1}^{J+1}C_{ij}\theta_j
\Biggr).
\end{equation}
For the special case where self-reports are perfect ($\varphi_1 =
\varphi_0 = 1$), equation (\ref{eq2.2}) reduces to the nonparametric likelihood
for interval censored observations given in \citet{b68}.

In most settings, including the WHI, it is of interest to evaluate the
association of a vector of covariates with respect to the time to event
of interest. Let $\mathbf{Z}$ denote the $P \times1$ vector of
explanatory variables with the corresponding $P \times1$ vector of
regression coefficients denoted by $\bolds{\beta}$. To incorporate
the effect of covariates, we assume the proportional hazards model,
$\lambda(t|\mathbf{Z}=\mathbf{z})=\lambda_0(t)e^{\mathbf
{z}'\bolds{\beta}}$, or, equivalently, $S(t|\mathbf
{Z}=\mathbf{z})=S_0(t)^{e^{\mathbf{z}'\bolds{\beta}}}$.

To derive the form of the log-likelihood based on the assumption of the
proportional hazards model, we\vspace*{1pt} first reparameterize the log likelihood
in (\ref{eq2.2}) in terms of the survival function, $\mathbf
{S}=(1=S_1,S_2,\ldots,S_{J+1})^T$, where $S_j=\operatorname{Pr}(X > \tau
_{j-1})$. Since $S_j = \sum_{l=j}^{J+1}{\theta_l}$, the\vspace*{2pt} vector of
interval probabilities can be expressed as $\bolds{\theta} = T_r
\mathbf{S}$, where $T_r$ is the $(J+1) \times(J+1)$ transformation
matrix. Let $C = [C_{ij}]$ denote the $N \times(J + 1)$ matrix of the
coefficients, $C_{ij}$, and let the $N \times(J+1)$ matrix $D$ be
defined as $D_{N \times(J+1)}=C\times T_r$.
Then, the log-likelihood function for the one-sample setting in (\ref{eq2.2})
can be expressed as
%
%e2.2 #&#
\begin{equation}
\label{survivallik} l(\mathbf{S})=\sum_{i=1}^N
\log\Biggl(\sum_{j=1}^{J+1}D_{ij}S_j
\Biggr),
\end{equation}
where $S_1=1$ and $S_2, S_3, \ldots, S_{J+1}$ are the
unknown parameters of interest.

Let $1=S_1>S_2>\cdots>S_{J+1}$ denote the baseline survival functions
(i.e., corresponding to $\mathbf{Z}=\mathbf{0}$), evaluated at
the left boundaries of the intervals $[0,\tau_1), [\tau_1,\tau_2),\ldots
,[\tau_J,\infty)$. Then, for subject $i$, with corresponding covariate
vector $\mathbf{z}_i$, $S_j^{(i)}=(S_j)^{e^{\mathbf
{z}_i'\bolds{\beta}}}$. Thus, the log-likelihood function for a
random sample of $N$ subjects can be expressed as
%
%e2.3 #&#
\begin{equation}
\label{loglikcov} l(\mathbf{S},\bolds{\beta})=\sum
_{i=1}^N\log\Biggl(\sum
_{j=1}^{J+1}D_{ij}(S_j)^{e^{\mathbf{z}_i'\bolds{\beta}}}
\Biggr).
\end{equation}

The elements of the $D$ matrix are functions of the observed data
including the visit times and corresponding self-reported results, as
well as the constants $\varphi_0, \varphi_1$. Assuming that $\varphi_0,
\varphi_1$ are known constants, the maximum likelihood estimates of the
unknown parameters $\beta_1,\ldots,\beta_{P}, S_2,\ldots,S_{J+1}$ can
be obtained by numerical maximization of the log-likelihood function,
subject to the constraints that $1>S_2>S_3>\cdots>S_{J+1}>0$.
Statistical inference regarding the parameters of interest ($\beta
_1,\ldots,\beta_{P}, S_2,\ldots,S_{J+1}$) can be made by using
asymptotic properties of the maximum likelihood estimators [\citet
{cox}]. The estimated covariance matrix of the maximum likelihood
estimates can be obtained by inverting the Hessian matrix. Hypothesis
tests regarding the unknown parameters can be carried out using the
likelihood ratio or Wald test.

%s2.2 #&#
\subsection{Misclassification at study entry}
In this section we incorporate the setting in which a self-report of
being event(disease)-free at baseline or study entry is used as the
inclusion criterion. The evaluation of the association between statin
use and risk of incident diabetes in the WHI was based on all women who
self-reported to be diabetes-free at baseline [\citet{ma2012}].
However, diabetes self-reports at study entry in the WHI have been
found to be less than perfect---the study by \citet{b46} found
that the negative predictive value of prevalent diabetes at baseline
was approximately 97\%, that is, 3\% of women who self-reported as
being diabetes-free were, in fact, diabetic. In this situation, the
assumption that $S(0)=1$ is invalid.

For the $i$th subject, let $G_i$ denote the baseline binary
self-report, where $G_i=1$ denotes a self-report indicating that the
event of interest has already occurred and $G_i=0$ denotes otherwise.
Similarly, let $B_i$ denote the true event status at baseline. In other
words, $B_i=1 \overset{\mathrm{def}}{=} X_i \le0 $ and $B_i=0 \overset{\mathrm{def}}{=}
X_i>0$. Consider a subject who has a negative self-report at baseline
(i.e., $G_i=0$) and is thus included in the data set. As before, let the
vector of observed self-reports for the $i$th subject be denoted by
$\mathbf R_i$. Let the negative predictive value of self-reports at
baseline be denoted by~$\eta$, that is, $\operatorname{Pr}(B_i=0|G_i=0) = \eta
$. Then the likelihood function for the $i$th subject can be expressed as
%
%e2.4 #&#
\begin{eqnarray}
L_i &=& \operatorname{Pr}(\mathbf R_i,
\mathbf{t}_i, n_i |G_i=0)\nonumber
\\
&=& \eta\operatorname{Pr}(\mathbf R_i, \mathbf{t}_i,
n_i|B_i=0,G_i=0)
\\
&&{} +(1-\eta) \operatorname{Pr}(\mathbf R_i, \mathbf{t}_i, n_i|B_i=1,G_i=0).\nonumber
\end{eqnarray}
We assume that subjects who self-report negative ($G_i = 0$) and are
truly negative for event at baseline ($B_i=0$) are a random sample from
all subjects who are true negative at baseline. Then we have $\operatorname{Pr}(\mathbf R_i, \mathbf{t}_i, n_i|B_i=0,G_i=0) =
\operatorname{Pr}(\mathbf R_i, \mathbf{t}_i, n_i|B_i=0) $, which corresponds to
the likelihood function derived in Section~\ref{s:likelihood}.\vspace*{-2pt} Thus,
$\operatorname{Pr}(\mathbf R_i, \mathbf{t}_i, n_i|B_i=0,G_i=0)=\sum
_{j=1}^{J+1}D_{ij}(S_j)^{e^{\mathbf{z}_i'\bolds{\beta}}}$.
Moreover, $\operatorname{Pr}(\mathbf R_i, \mathbf{t}_i,
n_i|B_i=1,G_i=0)=D_{i1}(S_1)^{e^{\mathbf{z}_i'\bolds{\beta}}}$.

The likelihood function for the $i$th subject has the form
\begin{eqnarray}\label{eq2.6}
\nonumber
L_i(\bolds{\beta}, \mathbf{S}) &=&\eta\sum
_{j=1}^{J+1}D_{ij}(S_j)^{e^{\mathbf{z}_i'\bolds{\beta}}}+(1-
\eta) D_{i1}(S_1)^{e^{\mathbf{z}_i'\bolds{\beta}}}
\nonumber\\[-8pt]\\[-8pt]\nonumber
&=&\sum_{j=1}^{J+1}D_{ij}'(S_j)^{e^{\mathbf{z}_i'\bolds{\beta}}},
\end{eqnarray}
where $D_{i1}'=D_{i1}$ and $D_{ij}' = \eta D_{ij}$ for $j>1$. Thus, the
likelihood function incorporating baseline misclassification has the
same general form as in equation (\ref{loglikcov}). The likelihood
function in equation~(\ref{loglikcov}) can be obtained as a special
case when $\eta=1$ in equation (\ref{eq2.6}).

%s2.3 #&#
\subsection{Time-varying covariates}
We consider the situation where covariate values can change with time
and are collected at each visit. Let $\mathbf{z}_{ij}$ denote the
$p \times1$ vector of covariate values for subject $i$ at time $\tau
_j$. In extending the likelihood function [equation (\ref{loglikcov})]
to handle time-varying covariates, we make the additional assumption
that the values of the covariates $\mathbf{z}_{ij}$ remain constant
during the interval $[\tau_j, \tau_{j+1})$. Let $\Lambda_j$ denote the
cumulative hazard function during the period of $[\tau_j, \tau_{j+1})$
for the subjects in the reference group (i.e., $\mathbf{Z}=0$).
Under the model $\lambda_{\mathbf{z}_{i}} (t) = \lambda_{\mathbf
{0}} (t) e^{\bolds{\beta} \mathbf{z}_{i}}$, the corresponding
cumulative hazard function during the period $[\tau_j, \tau_{j+1})$ for
subject $i$ is equal to $\Lambda_j\exp({\mathbf{z}_{ij}'\bolds{\beta}})$.
The survival function at $\tau_{j-1}$ can then be expressed as
\[
S_j^{(i)} = \exp\Biggl(-\sum_{j'=0}^{j-2}
\Lambda_{j'}\exp\bigl(\mathbf{z}_{ij'}'\bolds{\beta}
\bigr) \Biggr),
\]
where $j = 2, \ldots, J+1$, where $S_1^{(i)} = 1$. The log-likelihood
function can be expressed as a function of the derived $S_j^{(i)}$,
\[
l(\mathbf{S},\bolds{\beta})=\sum_{i=1}^N
\log\Biggl(\sum_{j=1}^{J+1}D_{ij}S_j^{(i)}
\Biggr).
\]
The log-likelihood function can be optimized with respect to the
parameters $\Lambda_0, \ldots, \Lambda_{J-1}$ and $\beta_1, \ldots,
\beta_P$ subject to constraints $\Lambda_j \ge0$. In practice, if a
subject has missing visits or missing covariate values at some visits,
one can carry forward the last observation as one approach to impute
missing covariate values. However, unless the proportion of missing is
very small, these ad hoc approaches toward handling missing data may
result in biased estimates of parameters and their associated standard errors.

%s2.4 #&#
\subsection{Unknown sensitivity and specificity}

Identifiability of the sensitivity and specificity parameters is
closely tied to the study design and the paradigm used for determining
number and timing of visits (tests). For example, in several
epidemiological cohorts in which self-reported outcomes of chronic
diseases such as diabetes are collected, data collection on the
incidence of the condition ceases following the first positive
self-report. In such study designs, it is implicitly assumed that
self-reports following the first positive self-report will be positive
with probability 1, thus, subsequent self-reports are noninformative.
In settings that incorporate an adaptive testing paradigm, the form of
the likelihood is shown in equation (\ref{loglikcov})---while this is a function of
the constants $\varphi_1, \varphi_0$ that characterize the sensitivity
and specificity of self-reports, these parameters cannot be estimated
jointly with the parameters of interest, namely, $\beta_1, \ldots, \beta
_p, S_2, \ldots, S_{J+1}$. If the sensitivity and specificity
parameters are unknown, an augmented study design in which a subset of
subjects are given a perfect diagnostic test in addition to
self-reported questionnaires could be considered. In these studies, the
parameters $\varphi_1, \varphi_0$ can be jointly estimated with the
unknown parameters of interest. A similar approach was proposed by
\citet{lyles2011} for mismeasured outcomes in logistic regression models.

In other clinical settings, the mismeasured outcome arises from
laboratory-based diagnostic tests characterized by imperfect
sensitivity and specificity. When the testing paradigm involves giving
the diagnostic test according to a predetermined testing schedule, the
form of the likelihood can be shown to be identical to that in equation
(2.4) [\citet{b6}]. In this case, it is possible to observe
seemingly inconsistent patterns of test results where one or more
negative test results follow a positive result. Examples include data
collected from DNA PCR assays to detect HIV infection in infants in
pediatric HIV clinical trials. Studies in which subjects are tested
according to a predetermined testing schedule, the sensitivity,
specificity parameters ($\varphi_1, \varphi_0$) can be jointly
estimated with the unknown parameters of interest [\citet{b49}].

%s3 #&#
\section{Simulation}
\label{s:simulation}
In this section we present results from simulation studies to
illustrate the effects of (1) error-prone self-reported outcomes; and
(2) misclassification at study entry. We present the effects of these
factors with regard to the bias associated with the estimated
regression parameter of interest.

%s3.1 #&#
\subsection{Effects of error-prone self-reported outcomes}\label{s:sim1}
We present average results from 1000 simulated data sets in which 1000
subjects were randomly assigned to two exposure groups with equal
proportion, assuming all subjects were event-free at baseline (i.e.,
$X_i > 0$ for all $i$). We assumed that there is a single binary
covariate of interest $Z_i$, corresponding to the exposure status of
the $i$th subject. The associated regression parameter in the
likelihood [equation (\ref{loglikcov})] was set to $\beta= 1$. For each
subject, self-reported questionnaires were collected at 8 scheduled
visits over a duration of 8 years, each with a random missing
probability of 30\%. All self-reports following the first positive
report were assumed to be positive with probability 1. The simulation
mechanism assumed that the time to the event of interest $X$ followed
an exponential distribution. The hazard rate $\lambda$ governing the
time to the event of interest in the reference group ($Z_i=0$) was set
to equal 0.0132 or 0.0866, corresponding to a cumulative incidence by
study end ($1-S_{J+1}$) of 0.10 or 0.50, respectively. As shown in
Table~\ref{tab01}, we compare results across several sets of values for
the parameters $(\varphi_1, \varphi_0)$, corresponding to the
sensitivity and specificity of self-reports.

%
%t1 #&#
\begin{table}
\tabcolsep=0pt
\caption{Comparing estimates of the regression parameter $\beta$ from
an ``adjusted'' analysis that accounts for the error in
self-reported outcomes to an ``unadjusted'' analysis that
incorrectly assumes that self-reports are perfect}\label{tab01}
\begin{tabular*}{\tablewidth}{@{\extracolsep{\fill}}@{}ld{1.3}ccd{3.1}ccd{2.1}@{}}
\hline
$\bolds{\varphi_1}$ & \multicolumn{1}{c}{$\bolds{\varphi_0}$} & \multicolumn{1}{c}{$\bolds{S_{J+1}}$} &
\multicolumn{1}{c}{\textbf{Analysis type}} & \multicolumn{1}{c}{\textbf{Bias (\%)}} &
\multicolumn{1}{c}{\textbf{Std Err}} & \multicolumn{1}{c}{\textbf{RMSE}} & \multicolumn{1}{c@{}}{\textbf{Coverage (\%)}}\\
\hline
0.75 & 1.00 & 0.90 & Adjusted & 0.3& 0.17 & 0.17 & 96.8 \\
0.75 & 1.00 & 0.90 & Unadjusted & 0.1 & 0.17 & 0.17 & 97.0 \\[3pt]
1.00 & 0.75 & 0.90 & Adjusted& -6.7 & 0.82 & 0.82 & 93.8 \\
1.00 & 0.75 & 0.90 & Unadjusted& -90.2 & 0.07 & 0.90 & 0.0 \\[3pt]
0.61 & 0.995 & 0.90 & Adjusted & 1.4 & 0.21 & 0.22 & 94.9 \\
0.61 & 0.995 & 0.90 & Unadjusted & -16.4 & 0.17 & 0.23 & 82.9 \\[3pt]
0.75 & 1.00 & 0.50 & Adjusted & 0.1 & 0.09 & 0.09 & 95.1 \\
0.75 & 1.00 & 0.50 & Unadjusted & -1.9 & 0.09 & 0.09 & 93.5 \\[3pt]
1.00 & 0.75 & 0.50 & Adjusted & 0.2 & 0.19& 0.19 & 94.4 \\
1.00 & 0.75 & 0.50 & Unadjusted & -59.2 & 0.07 & 0.60 & 0.0 \\[3pt]
0.61 & 0.995 & 0.50 & Adjusted & 0.5 & 0.09 & 0.09 & 94.2 \\
0.61 & 0.995 & 0.50 & Unadjusted & -6.9 & 0.08 & 0.11 & 86.7 \\
\hline
\end{tabular*}
\end{table}

In Table~\ref{tab01}, for each parameter setting, we present estimates
of bias, associated standard error, root mean square error (RMSE) and
coverage probability associated with the estimation of $\beta$.
Coverage probability was calculated as the proportion of data sets in
which the 95\% confidence interval for $\beta$ contains its true value.
We compare results from two sets of analyses for estimating $\beta$:
(a) maximizing the likelihood presented in equation (\ref{loglikcov}),
assuming that the true values of $\varphi_1, \varphi_0$ are known; and
(b) maximizing the likelihood presented in equation (\ref{loglikcov}),
assuming that self-reports are perfect (i.e., $\varphi_1=\varphi
_0=1$). In general, when the true values of $\varphi_0, \varphi_1$ are
incorporated into the analysis, the estimates of $\beta$ are nearly
unbiased. Similarly, the true coverage probability corresponding to a
95\% confidence interval is close to its nominal value. On the other
hand, when self-reports are incorrectly assumed to be perfect, the
estimates of $\beta$ may be significantly biased, especially in
settings where $\varphi_0$ is low. When $\varphi_0 \ll 1$, early false
positive self-reports result in significant loss of information due to
premature cessation of data collection. In this case, coverage
probabilities deviated significantly from 95\%. Last, incorporating the
uncertainty in error-prone self-reports increases the standard error of
the maximum likelihood estimates of $\beta$.

We note that while the true event times were simulated based on the
exponential distribution, the proposed methods make no distribution
assumptions. Thus, the performance of the proposed methods does not
depend on the underlying distributions of the event times. When event
times were simulated based on a Weibull distribution, similar results
were observed (results available upon request).

%s3.2 #&#
\subsection{Effects of misclassification at study entry}
In this simulation we incorporate the setting in which an error-prone,
self-report of being\break event(disease)-free at study entry is used as the
inclusion criterion. As before, let $\eta$ denote the negative
predictive value of the baseline self-report. That is, each subject
included in the study has a probability of $1-\eta$ of having already
experienced the event of interest prior to study entry. Each simulated
data set included 1000 subjects, of whom $1000 \times(1-\eta)$ had
already experienced the event of interest prior to entry into the study
(i.e., $X < 0$). The data were simulated as described in Section~\ref
{s:sim1}, where $\varphi_1=0.61$ and $\varphi_0=0.995$. We compare
results for various settings by varying the cumulative incidence of the
event of interest ($1-S_{J+1}$) to equal 0.10 or 0.50, and by varying
the value of $\eta$ to equal 0.99, 0.96 or 0.93.

%
%t2 #&#
\begin{table}[t]
\tabcolsep=0pt
\caption{Comparing estimates of the regression parameter $\beta$ from
an ``adjusted'' analysis that incorporates the possibility
of misclassification at baseline to an ``unadjusted'' analysis
that incorrectly assumes that all subjects are event-free at study
entry or that $\eta=1$. We assume that $\varphi_1= 0.61$ and $\varphi
_0= 0.995$}\label{tab02}
\begin{tabular*}{\tablewidth}{@{\extracolsep{\fill}}@{}lccd{2.1}ccc@{}}
\hline
$\bolds{S_{J+1}}$& $\bolds{\eta}$ & \multicolumn{1}{c}{\textbf{Analysis type}} &
\multicolumn{1}{c}{\textbf{Bias (\%)}} & \textbf{Std Err} & \textbf{RMSE} & \textbf{Coverage (\%)}\\
\hline
0.90 & 0.99 & Adjusted & 2.6 & 0.22 & 0.23 & 95.0 \\
0.90 & 0.99 & Unadjusted & -4.5 & 0.20 & 0.21 & 94.1 \\[3pt]
0.90 & 0.96 & Adjusted & 1.2 & 0.24 & 0.24 & 95.8 \\
0.90 & 0.96 & Unadjusted & -22.9 & 0.17 & 0.29 & 72.7 \\[3pt]
0.90 & 0.93 & Adjusted & 0.1 & 0.25 & 0.25 & 95.2 \\
0.90 & 0.93 & Unadjusted & -36.4 & 0.15 & 0.40 & 36.3 \\[3pt]
0.50 & 0.99 & Adjusted & 0.0 & 0.09 & 0.09 & 95.2 \\
0.50 & 0.99 & Unadjusted & -1.5 & 0.09 & 0.09 & 94.1 \\[3pt]
0.50 & 0.96 & Adjusted & 0.1 & 0.10 & 0.10 & 94.2 \\
0.50 & 0.96 & Unadjusted & -5.7 & 0.09 & 0.11 & 89.2 \\[3pt]
0.50 & 0.93 & Adjusted & 0.6 & 0.10 & 0.10 & 94.1 \\
0.50 & 0.93 & Unadjusted & -9.4 & 0.09 & 0.13 & 80.9 \\
\hline
\end{tabular*}
\end{table}

Table~\ref{tab02} presents the simulation results, averaged over 1000
data sets. We present results from an ``adjusted'' model
that properly accounts for misclassification at baseline based on the
likelihood presented in equation (\ref{eq2.6}) compared to the model in
equation (\ref{loglikcov}) that incorrectly assumes that $\eta=1$
(denoted ``unadjusted''). In both models, the true values
of the sensitivity and specificity are assumed. As expected, the
adjusted model is nearly unbiased and has uniformly lower bias when
compared to the unadjusted model. The bias of the unadjusted model
increases with decreasing values of negative predictive value ($\eta$),
and it is more pronounced when the cumulative incidence is low
($1-S_{J+1}=0.10$). In general, the inclusion of subjects who have
already experienced the event of interest at study entry results in the
exposure groups becoming less distinguishable. Thus, ignoring this
issue in data analysis results in estimates of exposure effects ($\beta
$) that are biased toward the null. In contrast, incorporating the
effect of baseline misclassification increases the standard error of
$\hat{\beta}$. The effects on the bias and the standard error of $\hat
{\beta}$ are reflected in the RMSE values---the adjusted model has
smaller RMSE than the unadjusted model in all settings except when
$S_{J+1}=0.9$ and $\eta=0.99$. The coverage probability of the adjusted
model is approximately 95\% in all settings considered in this study.
However, the coverage probability of the unadjusted model decreases
with decreasing negative predictive value ($\eta$) due to increased bias.

%s4 #&#
\section{Application: Risk of diabetes mellitus with statin use in the Women's Health Initiative}
\label{s:application}

\subsection*{Background}
We analyze data collected on 152,830 women from the
Women's Health Initiative (WHI) to evaluate the effects of statin use
on the risk of incident diabetes mellitus (DM). \citet{ma2012}
reported an increased risk of incident DM with baseline statin use
(multivariate-adjusted HR, 1.48; 95\% CI, 1.38--1.59). These results
were based on Cox proportional hazards models where the time to event
variable was calculated as the interval between enrollment date and the
earliest of the following: (1) date of annual medical history update
when new diabetes is self-reported (positive outcome); (2) date of last
annual medical update during which diabetes status can be ascertained
(censorship); or (3) date of death (censorship). The methods used in
\citet{ma2012} were based on the assumptions that: (1) all
subjects who self-reported as being diabetes-free at baseline were
truly not diabetic (i.e., $\eta=1$); and (2) the self-reports of
incident diabetes at each follow-up visit were error-free (i.e.,
$\varphi_1 = \varphi_0 = 1$). We compare the results from \citet
{ma2012} to results based on application of the likelihood-based
methods described in this paper.

\subsection*{Diabetes self-reports} Prevalent diabetes at baseline and
incident diabetes were assessed through self-reported questionnaires in
the WHI. At baseline and at each annual visit, participants were asked
whether she has ever received a physician diagnosis of and/or treatment
for diabetes when not pregnant since the time of the last self-report
(visit). Using data from a WHI substudy, estimates of sensitivity,
specificity and baseline negative predictive value of self-reported
diabetes outcomes were obtained by comparing self-reported outcomes to
fasting glucose levels and medication data [\citet{b46}]. A woman
was considered to be truly diabetic if she had either taken
anti-diabetic medication and/or had a fasting glucose level $\ge$126~mg/dl. From a representative subset of 5485 women, with information
at baseline on self-reported diabetes, fasting glucose levels and
medication inventory, we estimated that self-reports have a sensitivity
of 0.61, the specificity of 0.995, and a negative predictive value of
0.96 at baseline. These estimated parameter values are used in our
analysis. We used the following definitions: (1) sensitivity:
proportion of diabetics with a positive self-report; (2) specificity:
proportion of nondiabetics with a negative self-report; and (3)
negative predictive value: proportion of subjects who were
diabetes-free among those with a negative self-report. In practice,
estimating measurement error parameters from validation studies should
proceed with caution as validation studies may differ from their study
populations.

%
%t3 #&#
\begin{table}[b]
\tabcolsep=0pt
\caption{Analysis of the effects of statin use on incident diabetes
mellitus risk in the WHI}\label{tabwhi}
\begin{tabular*}{\tablewidth}{@{\extracolsep{\fill}}@{}lcccc@{}}
\hline
& & \textbf{Univariable/} & & \textbf{Hazard ratio}\\
\textbf{Statin variable type} & \textbf{Type of analysis} & \textbf{multivariable\tabnoteref{TT1}} & $\bolds{N}$ & \textbf{(95\% CI)}\\
\hline
Baseline statin & Proposed model & Univariable & 152,830 & 2.33~(2.12, 2.56) \\
Baseline statin & Proposed model & Multivariable& 138,338 & 1.81~(1.65, 1.99) \\[3pt]
Baseline statin & Cox model & Univariable & 152,830 & 1.69~(1.60, 1.78)\\
Baseline statin & Cox model & Multivariable & 138,338 & 1.54~(1.46, 1.63)
\\[6pt]
Time-varying statin & Proposed model & Univariable & 152,830 & 2.49~(2.31, 2.68) \\
Time-varying statin & Proposed model & Multivariable & 138,338 & 1.88~(1.75, 2.02) \\[3pt]
Time-varying statin & Cox model & Univariable & 152,830 & 1.65~(1.59, 1.72) \\
Time-varying statin & Cox model & Multivariable & 138,338 & 1.48~(1.42, 1.54) \\
\hline
\end{tabular*}
\tabnotetext{TT1}{Covariates adjusted include race, smoking status, alcohol intake, age,
education, WHI study, BMI, recreational physical activity, dietary energy
intake, family history of diabetes and hormone therapy use.}
\end{table}

\subsection*{Methods} The analysis data set included 152,830 women out of a
total of 161,808 women enrolled in the WHI. Women who self-reported
diabetes at baseline or those who ever took Cerivastatin were excluded.
In addition, women with missing data at baseline on diabetes status or
medication inventory were excluded [\citet{ma2012}]. The results
presented here are based on follow-up until 2010. The median duration
of follow-up was 12.1 years, including 1,688,967 person-years of total
follow-up. During the course of follow-up, 10.4\% of women
self-reported being diagnosed with diabetes. Information on statin use
was obtained from medical inventory information, which was available
for selected follow-up years. Information on statin use was available
for 152,830, 59,505, 128,507, 55,043 and 12,039 subjects at baseline, years
1, 3, 6 and 9, respectively. Models included either baseline statin use
or statin use as a time-varying covariate---in the latter case, the
most recent medication inventory data available was carried forward for
time points at which current medication use was not collected. In
multivariable models, other covariates included race, smoking status,
alcohol intake, age, education, WHI study, BMI, recreational physical
activity, dietary energy intake, family history of diabetes and hormone
therapy use [\citet{ma2012}]. We assumed that self-reports
following the first report of incident diabetes are noninformative.
Annual visit times were rounded to the nearest year in order to limit
the number of parameters estimated to describe the baseline survival
function ($S_2, \ldots, S_{J+1}$).

\subsection*{Results} Table~\ref{tabwhi} presents the estimated hazard ratio
(95\% confidence interval) for statin use by modeling statin use at
baseline or as a time-varying covariate. For each, we present results
from univariable models as well as multivariable models incorporating
potential confounders. In each setting, the results from the methods
proposed in this paper are compared to results from Cox models. In all
models, by incorporating the imperfect sensitivity and specificity of
self-reports and the potential misclassification at study entry, the
hazard ratio of statin use is consistently increased when comparing to
the corresponding Cox models. Using the proposed methods in equation
(\ref{eq2.6}), the hazard ratio for baseline statin use from univariate
analysis was 2.33 (95\% CI: 2.12--2.56). In the multivariable model, the
hazard ratio of baseline statin use was 1.81 (95\% CI: 1.65--1.99),
suggesting a relatively strong confounding effect. When statin use was
modeled as a time-varying covariate, the hazard ratios of statin use
from univariate and multivariate models were 2.49 (95\% CI: 2.31--2.68)
and 1.88 (95\% CI: 1.75--2.02), respectively.

The goodness of fit of the multivariable model incorporating statin use
as a time-varying covariate was assessed in an augmented model that
included 2 additional terms corresponding to the interactions of time
periods (in years) $(3,6]$ and $(6,16]$ with statin use. This model allows
the effect of statin use to vary between the time periods $(0,3]$, $(3,6]$
and $(6,16]$ years. The Wald test $p$ values corresponding to the
interactions of statin use with the time periods $(3,6]$ and $(6,16]$ were
0.89 and 0.11, respectively; these results indicate that the augmented
model provided no improvement in fit when compared to the model without
the additional interaction terms.

To evaluate how the results depend on the choice of parameters such as
sensitivity, specificity and baseline negative predictive value of
self-reported diabetes, we performed a sensitivity analysis by varying
each of these parameters. Table~\ref{tabsens} presents how the
estimated hazard ratio of statin use changes with different
combinations of the parameters. Statin use was modeled as a
time-varying covariate while simultaneously adjusting for potential
confounders. We observed that the estimated hazard ratio of statin use
is most sensitive to change in specificity. This is largely due to the
fact that the cumulative incidence of diabetes was low (10.4\%), and
thus false positive self-reports due to imperfect specificity have a
big influence on estimated parameters. In general, the hazard ratio of
statin use decreases as specificity increases. Changes in sensitivity
and negative predictive value at baseline have modest effects on the
resulting model fit.

%
%t4 #&#
\begin{table}
\tabcolsep=0pt
\caption{Statin use versus risk of incident diabetes mellitus in the
WHI---sensitivity analysis for varying sensitivity ($\varphi_1$),
specificity ($\varphi_0$) and baseline negative predictive value ($\eta
$) associated with diabetes self-reports. All models incorporate statin
use as a time-varying covariate and adjust for potential~confounders}\label{tabsens}
\begin{tabular*}{\tablewidth}{@{\extracolsep{\fill}}@{}lccc@{}}
\hline
& &  \textbf{Negative predictive} & \textbf{Hazard ratio}\\
\textbf{Sensitivity ($\bolds{\varphi_1}$)} & \textbf{Specificity ($\bolds{\varphi_0}$)} &  \textbf{value ($\bolds{\eta}$)} & \textbf{(95\% CI)}
\\
\hline
0.50 & 0.993 & 0.96 & 2.11~(1.92, 2.31) \\
0.50 & 0.993 & 0.98 & 2.10~(1.92, 2.30) \\[3pt]
0.50 & 0.995 & 0.96 & 1.93~(1.79, 2.08) \\
0.50 & 0.995 & 0.98 & 1.93~(1.79, 2.07) \\[3pt]
0.50 & 0.997 & 0.96 & 1.76~(1.65, 1.88) \\
0.50 & 0.997 & 0.98 & 1.77~(1.66, 1.88) \\[3pt]
0.61 & 0.993 & 0.96 & 2.05~(1.88, 2.24) \\
0.61 & 0.993 & 0.98 & 2.06~(1.89, 2.24) \\[3pt]
0.61 & 0.995 & 0.96 & 1.88~(1.75, 2.02) \\
0.61 & 0.995 & 0.98 & 1.89~(1.76, 2.03) \\[3pt]
0.61 & 0.997 & 0.96 & 1.73~(1.63, 1.84) \\
0.61 & 0.997 & 0.98 & 1.74~(1.64, 1.84) \\[3pt]
0.70 & 0.993 & 0.96 & 2.02~(1.85, 2.20) \\
0.70 & 0.993 & 0.98 & 2.03~(1.86, 2.21) \\[3pt]
0.70 & 0.995 & 0.96 & 1.86~(1.73, 2.00) \\
0.70 & 0.995 & 0.98 & 1.87~(1.74, 2.00) \\[3pt]
0.70 & 0.997 & 0.96 & 1.71~(1.61, 1.82) \\
0.70 & 0.997 & 0.98 & 1.72~(1.62, 1.82) \\
\hline
\end{tabular*}
\end{table}

The models presented here can be implemented using our freely available
R software package \textit{icensmis} [\citet{icensmis}] as described
in the supplemental material [\citet{suppA}].

%s5 #&#
\section{Discussion}\label{s:discussion}
Due to cost considerations, the use of self-reported outcomes is
common to diagnose prevalent and incident disease in large-scale
epidemiologic investigations. In this paper we present a
likelihood-based framework to model the association of a time-varying
covariate with a time to event outcome, that is observed through
periodically collected, error-prone, self-reported data. We incorporate
the possibility of erroneous inclusion of subjects who have already
experienced the event of interest prior to study entry as a result of
the use of self-reported outcomes at baseline in determining the study
population. R code for implementing the models proposed here are
presented in the supplemental material [\citet{suppA}].

We presented results from simulation studies to assess the impact of
ignoring error in self-reported outcomes---in all cases considered, the
use of statistical models that correctly accommodate the error inherent
in self-reports resulted in nearly unbiased estimates of the regression
parameter of interest. The largest bias as a result of ignoring error
in self-reported outcomes was found in settings where the cumulative
incidence was low and specificity was less than perfect. Models that
correctly accommodate error in self-reports also resulted in increased
variance of the estimated regression parameters. However, in most
settings, the RMSE values that combine the impact of bias and variance
of the estimated regression parameter favored the use of methods that
appropriately account for error in self-reported outcomes.

The methods proposed in this paper were applied to prospective data
from 152,830 women enrolled in the WHI to evaluate the effect of statin
use and risk of incident diabetes. By accounting for the imperfect
sensitivity, specificity and negative predictive value at baseline for
diabetes self-reports, we observed that the hazard ratio for statin use
was significantly larger than that estimated in naive analyses that
ignored the error in self-reported outcomes. In particular, the hazard
ratio of statin use in a multivariable model adjusted for potential
confounders was 1.88 (95\% CI: 1.75--2.02) as compared to the
multivariable hazard ratio estimate from Cox model 1.48 (95\% CI: 1.42--1.54).

In the methods developed here, we assumed that the sensitivity and
specificity of self-reported outcomes are invariant with respect to
time since entry and independent of covariates. In many real-world
settings, this assumption may result in over-simplified models,
particularly in applications in which visits are unequally spaced. In
addition, the methods developed here assumed that the parameters
governing the characteristics of self-reported outcomes are known.
However, in many cases these are estimated values---in this context, it
would be useful to extend the methods proposed here to consider study
designs including validation subsets that would allow joint estimation
of the sensitivity and specificity of self-reported outcomes together
with the other parameters of interest.

\section*{Acknowledgments}
We are grateful to the Editors and referees for their helpful comments.

\begin{supplement}[id=suppA]
%\sname{Supplement A}
\stitle{Tutorial for using the R package \textit{icensmis}}
\slink[doi]{10.1214/15-AOAS810SUPP} %[doi,text={...}] - jei reikia
%suskaldyti doi
\sdatatype{.pdf}
\sfilename{aoas810\_supp.pdf}
\sdescription{We present a short tutorial using the R package \textit{icensmis} to perform the analysis described in this paper.}
\end{supplement}

%\begin{appendix}
%\section{}
%\end{appendix}

% zodis "Acknowledgments" paliekamas pagal autoriu
%\

% imsref loaded by linak, 2015-03-30 10:13:32
%

\printaddresses
\end{document}